\documentclass[a4paper,11pt]{article}
\usepackage{pos}

\renewcommand{\logo}{\relax}

\usepackage{caption}
\usepackage{subcaption}

\DeclareMathOperator{\Tr}{Tr}

\title{One-loop RG flow for adjoint multiscalar gauge theory}

\author*{Nadia Flodgren}

\affiliation{The Oskar Klein Centre \& Department of Physics, Stockholm University,\\
  AlbaNova, 106 91 Stockholm, Sweden}

\emailAdd{nadia.flodgren@fysik.su.se}

\abstract{We study the one-loop renormalisation of 4d SU(N) Yang-Mills theory with $M$ adjoint representation scalar multiplets. We calculate the coupled one-loop renormalization group flows for this theory by developing an algebraic description, which we find to be characterised by a non-associative algebra of marginal couplings. The 4d one-loop beta function of the gauge coupling $g^2$ vanishes for the case $M = 22$, which is intriguing for string theory. There are real fixed flows (fixed points of $\lambda/g^2$) only for $M\geq406$, rendering one-loop fixed points of the gauge coupling and scalar couplings incompatible.}

\FullConference{The European Physical Society Conference on High Energy Physics (EPS-HEP2023)\\
 21-25 August 2023\\
Hamburg, Germany\\}

\begin{document}
\maketitle

\section{Introduction}

Gauge interactions have been argued to be necessary for weakly coupled UV fixed points in 4D QFTs \cite{Coleman:1973sx, Bond:2018oco}. We study the one-loop RG flow of multiscalar gauge theory in 4D at large N with massless adjoint scalars that interact quartically. The adjoint scalar gauge theories are of interest due to their use in describing the dynamics on branes in $D3-$brane theories in string theory. In our model the adjoint scalar multiplets are invariant under $O(M)$ symmetry and the gauge group is $SU(N)$. The quartic interaction term is
\begin{equation}
\mathcal{L}_{int} = -\frac{1}{4!}\lambda_{\bar{A}\bar{B}\bar{C}\bar{D}}\phi_{\bar{A}}\phi_{\bar{B}}\phi_{\bar{C}}\phi_{\bar{D}}.
\end{equation}
The multi-index $\bar{A}=aA$ where $a=1,\dots,M$ is the scalar multiplet index and $A=1,\dots,N^2-1$ is the Lie algebra index.

This presentation is based on my work with Bo Sundborg \cite{Flodgren:2023lyl}.

\section{Algebraic description of one-loop RG flow of adjoint multiscalar gauge theory}

The general one-loop beta functions for a multiscalar gauge theory are \cite{Machacek:1984zw, Luo:2002ti} 
\begin{equation} \label{eq:beta}
\begin{split}
&\mu \frac{d}{d\mu}\lambda_{\bar{A}\bar{B}\bar{C}\bar{D}}
=\beta_{\bar{A}\bar{B}\bar{C}\bar{D}} 
=\frac{1}{(4\pi)^2} \bigg( \Lambda_{\bar{A}\bar{B}\bar{C}\bar{D}}^2-3g^2\Lambda^S_{\bar{A}\bar{B}\bar{C}\bar{D}} +3g^4A_{\bar{A}\bar{B}\bar{C}\bar{D}} \bigg),
\end{split}
\end{equation}
where
\begin{equation} \label{eq_LLA}
\begin{split}
&\Lambda^2_{\bar{A}\bar{B}\bar{C}\bar{D}}=\frac{1}{8}\sum_{\text{perms}} \lambda_{\bar{A}\bar{B}\bar{E}\bar{F}}\lambda_{\bar{E}\bar{F}\bar{C}\bar{D}} \\
&\Lambda^S_{\bar{A}\bar{B}\bar{C}\bar{D}}=\sum_i C(i)\lambda_{\bar{A}\bar{B}\bar{C}\bar{D}} \\
&A_{\bar{A}\bar{B}\bar{C}\bar{D}} =\frac{1}{8}\sum_{\text{perms}} \{ \theta^E,\theta^F\}_{\bar{A}\bar{B}}\{ \theta^E,\theta^F\}_{\bar{C}\bar{D}}.
\end{split}
\end{equation}
The gauge coupling is called $g$, the sum over permutations is over the permutations of the multi-indices, the sum over $i$ is over the external legs, $C(i)$ is the quadratic Casimir $C_2(G)$ for the gauge group $G$ and $\theta_{\bar{A}\bar{B}}^C$ is a reducible representation of the gauge Lie algebra. 

The one-loop beta functions for multiscalar gauge theory can be characterized by an algebra, as described in more detail in our work \cite{Flodgren:2023lyl}. 
We consider only marginal quartic couplings $\lambda_{\bar{A}\bar{B}\bar{C}\bar{D}}$ and can think of the couplings as a vector space. For a purely scalar theory the one-loop beta function $\beta_{\lambda}$ is quadratic in $\lambda$, as is seen from the scalar-scalar interaction term in (\ref{eq_LLA}). This allows us to define a product of the marginal couplings
\begin{equation} \label{eq_d0}
\begin{split}
\lambda \diamond \kappa &\equiv 
\frac{(4\pi)^2}{2}(\beta_{\lambda+\kappa}-\beta_{\lambda}-\beta_{\kappa}).
\end{split}
\end{equation}
Here we ignored the gauge terms in the beta functions since they do not affect the product. The product is commutative but not generally associative.

It is useful to express the algebra using a basis of symmetric rank four tensor structures  $g^{k}_{\bar{A}\bar{B}\bar{C}\bar{D}}$ where $\lambda_{\bar{A}\bar{B}\bar{C}\bar{D}}=\lambda_kg^{k}_{\bar{A}\bar{B}\bar{C}\bar{D}}$. The basis elements $g^{k}_{\bar{A}\bar{B}\bar{C}\bar{D}}$ are the tensor structures of the marginal operators of the theory, i.e. the quartic invariants (indexed by $k$). 
The beta function is $\beta_{\bar{A}\bar{B}\bar{C}\bar{D}}=\beta_kg^k_{\bar{A}\bar{B}\bar{C}\bar{D}}$ where $\beta_k = \frac{d\lambda_k}{d\ln(\mu)}$. We define the product   
\begin{equation} \label{eq_diamond}
\begin{split}
\left(g^m \diamond g^n\right)_{\bar{A}\bar{B}\bar{C}\bar{D}} = \frac{1}{8}\sum_{\text{perms}} g^m_{\bar{A}\bar{B}\bar{E}\bar{F}}g^n_{\bar{E}\bar{F}\bar{C}\bar{D}}  
  = C^{mn}_kg^k_{\bar{A}\bar{B}\bar{C}\bar{D}}.
\end{split}
\end{equation}
The coefficients $C^{mn}_k$ are the structure constants of the algebra. The product is known as the $\vee$-product in \cite{Michel:1971th}. The beta function is 
\begin{equation} \label{eq_bk1}
\begin{split}
(4\pi)^2 \beta_k = ( \lambda_m\lambda_n  C^{mn}_k - 3g^2C_1 \lambda_k  + 3g^4d_k),\\
\end{split}
\end{equation}
where $d_k$ are the coefficients of the gauge induced term and $C_1=4C_2(G)$. 

Let us study the specific example of our $SU(N)\times O(M)$ model, which has four quartic invariants indexed by $k=1s,1t,2s,2t$. They correspond to the symmetric tensor structures via\footnote{Note that we will from now on suppress the multi-indices when writing the basis elements $g^k_{\bar{A} \bar{B} \bar{C} \bar{D}} \rightarrow g^k$.}
\begin{equation} \label{eq_poly1}
\begin{aligned}
 \frac{1}{4 !} g_{\bar{A} \bar{B} \bar{C} \bar{D}}^{1 s} \phi_{\bar{A}} \phi_{\bar{B}} \phi_{\bar{C}} \phi_{\bar{D}}&=\frac{1}{2} \Tr \Phi_a \Phi_a \Phi_b \Phi_b \\
\frac{1}{4 !} g_{\bar{A} \bar{B} \bar{C} \bar{D}}^{1 t} \phi_{\bar{A}} \phi_{\bar{B}} \phi_{\bar{C}} \phi_{\bar{D}}&=\frac{1}{4} \Tr \Phi_a \Phi_b \Phi_a \Phi_b \\
\frac{1}{4 !} g_{\bar{A} \bar{B} \bar{C} \bar{D}}^{2 s} \phi_{\bar{A}} \phi_{\bar{B}} \phi_{\bar{C}} \phi_{\bar{D}}&=\frac{1}{2} \Tr \Phi_a \Phi_a \Tr \Phi_b \Phi_b  \\
 \frac{1}{4 !} g_{\bar{A} \bar{B} \bar{C} \bar{D}}^{2 t} \phi_{\bar{A}} \phi_{\bar{B}} \phi_{\bar{C}} \phi_{\bar{D}}&=\Tr \Phi_a \Phi_b \Tr \Phi_a \Phi_b,
\end{aligned}
\end{equation}
where $\Phi_a=\phi_{\bar{A}}T_A$ belongs to the adjoint representation and $T_A$ is matrix in the fundamental representation\footnote{We use the normalization $\Tr(T_AT_B)=\frac{1}{2}\delta_{AB}$ of the fundamental representation matrices $T_A$.}. The superscripts $s$ and $t$ stand for scalar and tensor product respectively, and $1$ and $2$ stand for single-trace and double-trace respectively. 

To take the large $N$ limit of the algebra for $g^k$ we must first rescale the couplings $\lambda_k$ in order to keep the 't Hooft couplings $\lambda_K$ constant: $\lambda_{1S}=N\lambda_{1s}$, $\lambda_{1T}=N\lambda_{1t}$ and $\lambda_{2S}=N^2\lambda_{2s}$, $\lambda_{2T}=N^2\lambda_{2t}$. The basis elements $g^k$ are rescaled to $g^K$ accordingly to keep $\lambda_kg^k=\lambda_Kg^K$ constant. Taking the large $N$ limit and dropping sub-leading terms simplifies the algebra. The algebra at large $N$ for the basis $g^K$ is 
\begin{equation}\label{eq_g}
\begin{aligned}
g^{1 S} \diamond g^{1 S} & =\frac{1}{2}(M+3) g^{2 S}+\frac{1}{2} g^{2 T}+\frac{1}{2}(M+3) g^{1 S} \\
g^{1 T} \diamond g^{1 T} & =\frac{1}{8}(M+2) g^{2 T}+\frac{1}{2} g^{1 S} \\
g^{1 S} \diamond g^{1 T} & =\frac{1}{2} g^{2 S}+\frac{1}{2} g^{2 T}+\frac{1}{2} g^{1 S}+g^{1 T} \\
g^{2 S} \diamond g^{2 S} & =M g^{2 S} \\
g^{2 T} \diamond g^{2 T} & =2 g^{2 T} \\
g^{2 S} \diamond g^{2 T} & =2 g^{2 S} \\
g^{1 S} \diamond g^{2 S} & =(M+1) g^{2 S} \\
g^{1 S} \diamond g^{2 T} & =2 g^{2 S}+g^{2 T} \\
g^{1 T} \diamond g^{2 S} & =g^{2 S} \\
g^{1 T} \diamond g^{2 T} & =g^{2 T} .
\end{aligned}
\end{equation}

The large $N$ algebra has several closed sub-algebras, $\{g^{2S}\},\{g^{2T}\},\{g^{2S},g^{2T}\},\{g^{1S},g^{2S},g^{2T}\}$, and two ideals $\{g^{2S}\},\{g^{2S},g^{2T}\}$. The brackets $\{\}$ denote a linear space with the $\diamond$-product that is spanned by the elements in the brackets. The algebra indicates which couplings induce running in other couplings. For example, the closed sub-algebra $\{g^{2T}\}$ indicates that a theory with the only non-zero coupling\footnote{Note that the gauge coupling or its coefficients $d_K$ in $\beta_K$ must also be vanishing here.} being $\lambda_{2T}$ is renormalizable in this limit. 
The ideals are closed sub-algebras that are stable under perturbations of elements outside the ideal. The sub-algebra $\{g^{2S}\}$ is an ideal because all products with $g^{2S}$ in (\ref{eq_g}) result only in the element itself. Physically the beta functions of the quotient algebra of an ideal are independent of the couplings in the ideal, giving the RG equations a hierarchical structure.

\section{Results: one-loop RG flow for adjoint multiscalar gauge theory}

Let us take a look at the beta functions for our model.
The one-loop beta function for the gauge coupling vanishes for $M=22$ adjoint scalars
\begin{equation}
\begin{aligned}
\beta_g & =-\frac{g^3}{16 \pi^2} \frac{22-M}{6} C_2(G). \\
\end{aligned}
\end{equation}
The RG flow is asymptotically UV free for $M<22$ and IR free for $M>22$. The critical value of $M=22$ has been noted before in \cite{Curtright:1981wv}. The critical value intrigued us as it could indicate a string theory connection since $22$ adjoint scalars in 4D would correspond to a low energy limit of $N$ parallel $D3$-branes in $26=22+4$ dimensions (the critical dimension of bosonic strings).

From the algebra (\ref{eq_g}) we calculate the scalar coupling beta functions, apart from the gauge induced terms which are calculated separately. We find no real fixed points for $M=22$ scalars. In order to relate to possible fixed points for a running gauge coupling we calculate the beta functions for the relative couplings $\mu_K=\lambda_K/g^2$. A fixed point for a relative coupling is called a fixed flow. The beta functions for the relative couplings at large $N$ are 
\begin{equation} \label{eq_beta}
\begin{aligned}
\beta_{1 S}(\mu)= & \frac{N g^2}{16 \pi^2}\left(\frac{M+3}{2} \mu_{1 S} \mu_{1 S}+\frac{1}{2} \mu_{1 T} \mu_{1 T}+\mu_{1 T} \mu_{1 S}-\left(\frac{M-22}{3}+12\right) \mu_{1 S}+6\right) \\
\beta_{1 T}(\mu)= & \frac{N g^2}{16 \pi^2}\left(2 \mu_{1 T} \mu_{1 S}-\left(\frac{M-22}{3}+12\right) \mu_{1 T}\right) \\
\beta_{2 S}(\mu)= & \frac{N g^2}{16 \pi^2}\left(M \mu_{2 S} \mu_{2 S}+\frac{(M+3)}{2} \mu_{1 S} \mu_{1 S}+4 \mu_{2 S} \mu_{2 T}+2 \mu_{2 S} \mu_{1 T}+2(M+1) \mu_{2 S} \mu_{1 S}\right. \\
& \left.+4 \mu_{2 T} \mu_{1 S}+\mu_{1 T} \mu_{1 S}-\left(\frac{M-22}{3}+12\right) \mu_{2 S}+6\right) \\
\beta_{2 T}(\mu)= & \frac{N g^2}{16 \pi^2}\left(2 \mu_{2 T} \mu_{2 T}+\frac{1}{2} \mu_{1 S} \mu_{1 S}+\frac{M+2}{8} \mu_{1 T} \mu_{1 T}+2 \mu_{2 T} \mu_{1 T}+2 \mu_{2 T} \mu_{1 S}+\mu_{1 T} \mu_{1 S}\right. \\
& \left.-\left(\frac{M-22}{3}+12\right) \mu_{2 T}+6\right).
\end{aligned}
\end{equation}
Note that these beta functions are dependent on $M$ but independent of the flow of the gauge coupling, meaning we can treat $g$ as a constant in searching for fixed points of the relative couplings. 

In the beta functions (\ref{eq_beta}) we can observe the implication of the quotient algebra of the double-trace ideal $\{g^{2S},g^{2T}\}$, i.e. that $\beta_{1S}$ and $\beta_{1T}$ are independent of the double-trace couplings $\lambda_{2S}, \lambda_{2T}$, and the implication of the ideal $\{g^{2S}\}$, which is that only $\beta_{2S}$ depends on $\lambda_{2S}$. 

\begin{figure}
     \centering
     \begin{subfigure}[b]{0.4\textwidth}
         \centering
         \includegraphics[width=\textwidth]{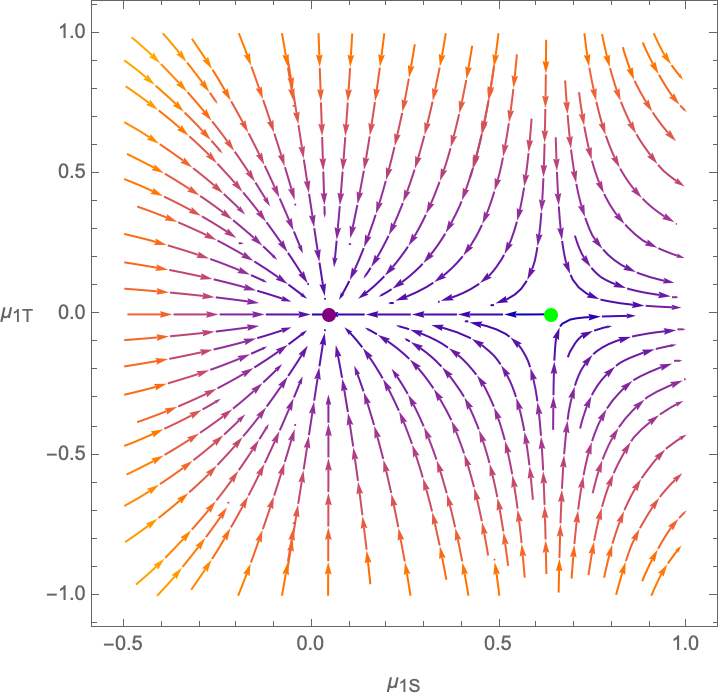}
         \caption{Single-trace RG flow for $M=425$.}
         \label{fig:ST}
     \end{subfigure}
     \hfill
     \begin{subfigure}[b]{0.4\textwidth}
         \centering
         \includegraphics[width=\textwidth]{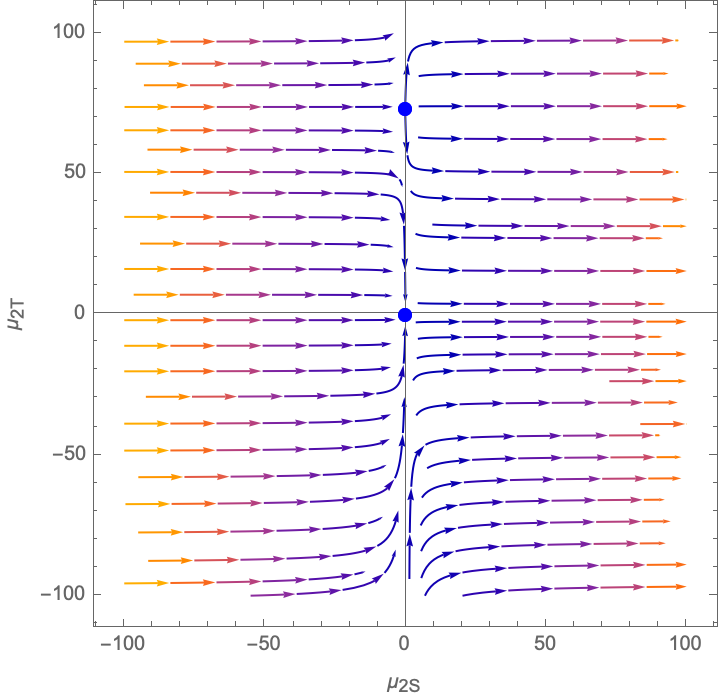}
         \caption{Double-trace RG flow for $M=425$.}
         \label{fig:DTtotal}
     \end{subfigure}
     \hfill
        \caption{The RG flow for $M=425$ in the single-trace space \ref{fig:ST} and double-trace space \ref{fig:DTtotal}. The single-trace space has one UV stable fixed flow (purple dot) and one of mixed stability (green dot). The double-trace space has four fixed flows (two for each blue dot) which are shown in Figure \ref{fig:RGtotal2}.}
        \label{fig:RGtotal}
\end{figure}

\begin{figure}
     \centering
     \begin{subfigure}[b]{0.4\textwidth}
         \centering
         \includegraphics[width=\textwidth]{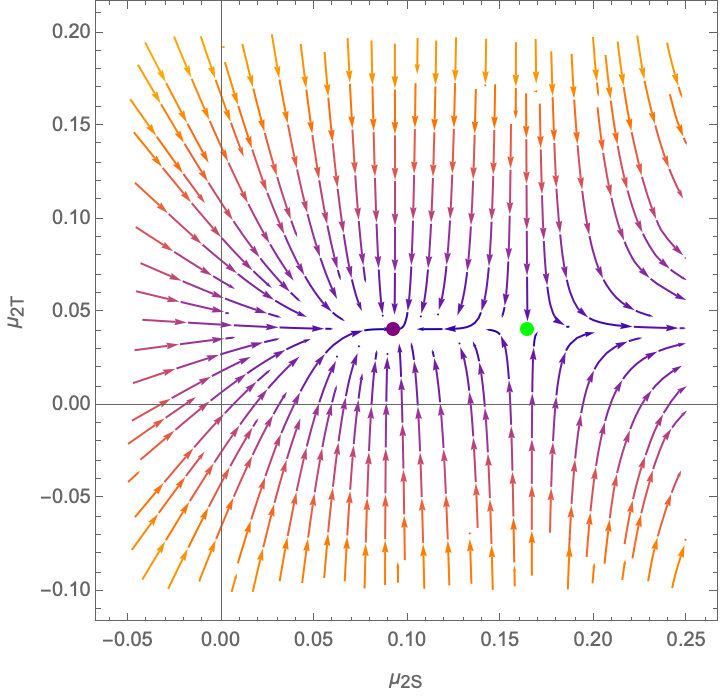}
         \caption{Double-trace RG flow for $M=425$, lower feature.}
         \label{fig:DT1}
     \end{subfigure}
     \hfill
     \begin{subfigure}[b]{0.4\textwidth}
         \centering
         \includegraphics[width=\textwidth]{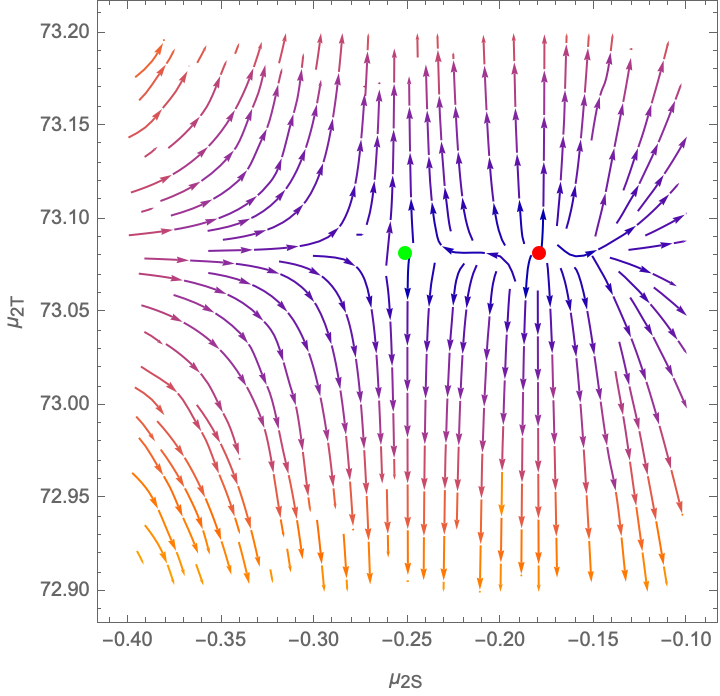}
         \caption{Double-trace RG flow for $M=425$, upper feature.}
         \label{fig:DT2}
     \end{subfigure}
     \hfill
        \caption{Double-trace RG flow for $M=425$. The lower feature has one UV stable fixed flow and one of mixed stability. The upper feature has one IR stable fixed flow (red dot) and one of mixed stability. Together the four fixed flows form a parallelogram which no flow can escape.}
        \label{fig:RGtotal2}
\end{figure}

In the space of single-trace couplings $(\lambda_{1S},\lambda_{1T})$ there are real fixed flows only for $M\geq 82$, for which there exists one UV-stable and one mixed stability fixed flow, see Figure \ref{fig:ST}.
The RG flow of the double-trace coupling space $(\lambda_{2S},\lambda_{2T})$ is seen in Figure \ref{fig:DTtotal} and \ref{fig:RGtotal2}. In this space there are only real fixed flows for $M\geq 406$. In the range $406 \leq M \leq 427$ there are four real fixed flows (one IR stable, one UV stable and two of mixed stability) and for $M\geq 428$ there are eight real fixed flows. 

In conclusion, the RG flow of the complete space of marginal quartic couplings only has real fixed flows for $M\geq 406$, for which the gauge coupling is IR free. Therefore, no complete gauge coupling and scalar coupling fixed point exists for the adjoint multiscalar gauge theory at large $N$ to one-loop order. 

\section{Outlook}

So far we have only studied the regular large $N$ limit but there exists other possible large $N$ limits we want to investigate, such as a Veneziano-like limit where both $N$ and $M$ are large. We are interested in seeing whether the algebra itself can tell us which limits have interesting RG flows.  

Concerning the specific $SU(N)\times O(M)$ model we studied, one could consider the case $M=22$ but for a complex CFT and look for stable complex CFTs. The stability of the potential, in the sense of a theory bounded from below, at the fixed points would then need to be checked. 

The main unanswered question we have is if the algebraic description of the RG flow can be adapter to higher loop orders. Related to this, we have also thought about studying the RG flow of models with different fields and interactions via the algebra, to see if the algebra can give new insights. Generalizing out method should be fairly simple since it relies on very general known one-loop equations.

\bibliography{BibliographyArticles}{}
\bibliographystyle{JHEP}

\end{document}